\documentclass[aps,prl,twocolumn,groupedaddress,superscriptaddress]{revtex4-2}
\usepackage{graphicx}
\usepackage{amsmath}
\usepackage{epstopdf}
\usepackage{float}
\usepackage{color}
\usepackage{appendix}
\usepackage{ulem}

\usepackage{multirow}
\usepackage{booktabs}
\usepackage{longtable}
\usepackage{array}     
\usepackage{tabularx}
\usepackage{hyperref}

\begin{document}
\title{Precision Measurement of Spin-Dependent Dipolar Splitting \\in $^6$Li \textit{p}-Wave Feshbach Resonances}
 

\author{Shuai Peng}
\thanks{These authors contributed equally to this work.}
\affiliation{School of Physics and Astronomy, Sun Yat-sen University, Zhuhai, Guangdong, China 519082}

\author{Sijia Peng}
\thanks{These authors contributed equally to this work.}
\affiliation{School of Physics and Astronomy, Sun Yat-sen University, Zhuhai, Guangdong, China 519082}

\author{Lijun Ren}
\thanks{These authors contributed equally to this work.}
\affiliation{School of Physics and Astronomy, Sun Yat-sen University, Zhuhai, Guangdong, China 519082}

\author{Shaokun Liu}
\affiliation{School of Physics and Astronomy, Sun Yat-sen University, Zhuhai, Guangdong, China 519082}

\author{Bin Liu}
\affiliation{School of Physics and Astronomy, Sun Yat-sen University, Zhuhai, Guangdong, China 519082}

\author{Jiaming Li}
\email[]{lijiam29@mail.sysu.edu.cn}
\affiliation{School of Physics and Astronomy, Sun Yat-sen University, Zhuhai, Guangdong, China 519082}
\affiliation{Guangdong Provincial Key Laboratory of Quantum Metrology and Sensing $\&$ School of Physics and Astronomy,	Sun Yat-Sen University, Zhuhai, China 519082}
\affiliation{Center of Quantum Information Technology, Shenzhen Research Institute of Sun Yat-sen University, Shenzhen, Guangdong, China 518087}
\affiliation{State Key Laboratory of Optoelectronic Materials and Technologies, Sun Yat-sen University, Guangzhou, China 510275}

\author{Le Luo}
\email[]{luole5@mail.sysu.edu.cn}
\affiliation{School of Physics and Astronomy, Sun Yat-sen University, Zhuhai, Guangdong, China 519082}
\affiliation{Guangdong Provincial Key Laboratory of Quantum Metrology and Sensing $\&$ School of Physics and Astronomy,	Sun Yat-Sen University, Zhuhai, China 519082}
\affiliation{Center of Quantum Information Technology, Shenzhen Research Institute of Sun Yat-sen University, Shenzhen, Guangdong, China 518087}
\affiliation{State Key Laboratory of Optoelectronic Materials and Technologies, Sun Yat-sen University, Guangzhou, China 510275}



\date{\today}
\begin{abstract}
The magnetic dipolar splitting of a \textit{p}-wave Feshbach resonance is governed by the spin–orbital configuration of the valence electrons in the triplet molecular state. We perform high-resolution trap-loss spectroscopy on ultracold $^{6}$Li atoms to resolve this splitting with sub-milligauss precision. By comparing spin-polarized ($|m_S| = 1$) and spin-mixture ($m_S = 0$) configurations of the triplet state, we observe a clear spin-dependent reversal in the splitting structure, confirmed via momentum-resolved absorption imaging. This behavior directly reflects the interplay between electron spin projection $m_S$ and orbital angular momentum $m_\ell$ in the molecular states. Our results provide a stringent benchmark for dipole–dipole interaction models and lay the groundwork for controlling  the pairing in \textit{p}-wave superfluid systems.

\end{abstract}
\maketitle

\section{INTRODUCTION}
Ultracold atomic gases near a \textit{p}-wave Feshbach resonance (FR) provide access to anisotropic interactions with nonzero orbital angular momentum, which are absent in conventional \textit{s}-wave resonances~\cite{Regal2003PRL90.053201, Bohn2000PRA61.053409, Gaebler2007PhysRevLett.98.200403, Chin2010Rev.Mod.Phys.82.1225-1286,Zhang2004PhysRevA.70.030702, Schunck2005Phys.Rev.A71.045601,Peng2024Phys.Rev.A110.L051301}. These anisotropic interactions give rise to novel quantum correlations and exotic pairing mechanisms, which underlie proposals for topological superfluidity~\cite{Gurarie2005Phys.Rev.Lett.94.230403, Cheng2005Phys.Rev.Lett.95.070404, Gurarie2007Ann.Phys.322.2-119, Suzuki2008Phys.Rev.A77.043629,Tewari2007PRL98.010506} and spin-polarized fermionic phases~\cite{You1999PRA60.2324, Girardeau2008Phys.Rev.Lett.100.200403,Koscik2023Phys.Rev.Lett.130.253401}.

A key feature of \textit{p}-wave FRs is the magnetic dipole--dipole interaction between the electronic spins of the atoms in the closed-channel molecular state~\cite{Explainreff}. This interaction is anisotropic and lifts the degeneracy of the orbital angular momentum projections ($m_\ell = 0, \pm1$), resulting in a characteristic dipolar splitting of the resonance into multiple components~\cite{Ticknor2004Phys.Rev.A69.042712}.

Such dipolar splitting serves as a distinct fingerprint of higher partial-wave resonances and is governed by the spin–orbital configuration of the valence electrons in the triplet molecular state. It provides a direct probe of magnetic dipole–dipole interactions and reveals detailed information about the internal structure of the molecular state. These features make $p$-wave FRs an especially sensitive testbed for coupled-channel (CC) theory and other high-precision quantum scattering models~\cite{Weiner1999RMP71.1,Gao1998PRA58.4222, Gao2005Phys.Rev.A72.042719,Li2018Phys.Rev.Lett.120.193402, Horn2024SA10.eadi6462,Lysebo2009PRA79.062704}.

\begin{figure}[htbp]
	\begin{center}
		\includegraphics[width=0.8\columnwidth]{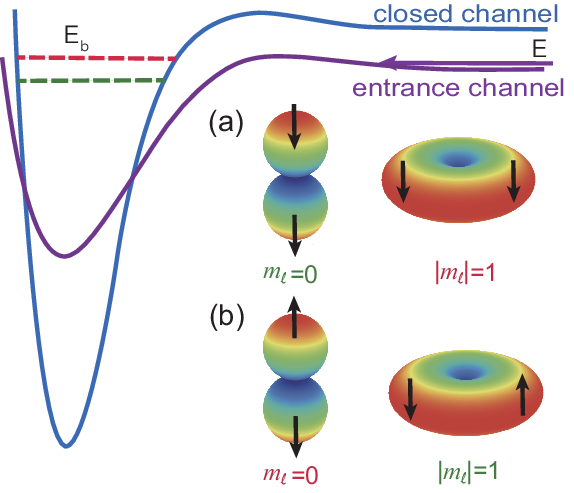}
	\end{center}
	\caption{Schematic of spin-dependent dipole-dipole interactions in $L=1$ orbitals. (a) In spin-polarized molecular states, the $m_\ell=0$ configuration, predominantly corresponding to head-to-tail alignment, alternates between attraction and repulsion, while $|m_\ell| = 1$ remain repulsive as its side-by-side motion. (b) In spin-mixtures molecular states, the $m_\ell=0$ configuration alternates, while $|m_\ell| = 1$ configuration become attractive, reversing dipolar splitting. A detailed comparison for $^6$Li \textit{p}-wave FRs, including CC bound-state calculations and the corresponding molecular states with atomic $m_s$, is provided in Supplementary Material (SM) Section~I.} 
	\label{physics}
\end{figure}

As illustrated in Fig.\ref{physics}, the spatial orientation of magnetic dipoles governs the anisotropy of the dipole--dipole interaction. In a $p$-wave molecular state formed by two atoms with polarized electron spins ($m_{s1}=m_{s2}=-1/2$) [Fig.~\ref{physics}(a)], the $|m_\ell| = 1$ orbital geometry corresponds to side-by-side motion, which leads to repulsive interactions. And the $m_\ell = 0$ configuration, predominantly corresponding to head-to-tail alignment, supports an angularly dependent potential with alternating attractive and repulsive regions. Thus, in spin-polarized systems such as $^{6}$Li $p$-wave FRs, the $m_\ell = 0$ molecular state lies at lower energy than the $|m_\ell| = 1$ states.

In contrast, when the molecular state is formed by atoms with mixed electron spins ($m_{s1}=1/2, m_{s2}=-1/2$), the resulting dipole orientations are  non-parallel [Fig.~\ref{physics}(b)]. This modifies the angular structure of the dipole--dipole interaction: side-by-side configurations ($|m_\ell| = 1$) can become attractive, while head-to-tail alignment ($m_\ell = 0$) instead exhibit alternating  of attraction and repulsion. Consequently, the dipolar splitting reverses polarity, and the $m_\ell = 0$ state appears at higher energy than the $|m_\ell| = 1$ components—signaling a qualitatively different interaction geometry in spin-mixture systems.

Previous experiments have successfully resolved dipolar splittings in spin-polarized systems [Fig.\ref{physics}(a)], including $^{40}$K near 198 G \cite{Ticknor2004Phys.Rev.A69.042712, Gunter2005PRL95.230401, Luciuk2016NP12.599} and $^{6}$Li near 159 and 215 G\cite{Gerken2019Phys.Rev.A100.050701}. In these cases, magnetic dipole--dipole interactions produce relatively large energy splittings—typically tens to hundreds of milligauss—which are readily resolved using conventional spectroscopy without requiring sub-milligauss resolution or extreme magnetic field stability.

However, dipolar splittings  in spin-mixture systems [Fig.~\ref{physics}(b)] have not been directly observed due to their weaker magnitude, making the corresponding doublet structure significantly harder to resolve. For instance, CC calculations predict that in the $p$-wave FR of a two-component $^6$Li Fermi gas $|1\rangle \equiv |F=1/2, m_F=1/2\rangle$ and $|2\rangle \equiv |F=1/2, m_F=-1/2\rangle$ near 185 G, the $m_\ell = 0$ resonance lies approximately $3.6$ mG below the $|m_\ell| = 1$ components~\cite{Chevy2005Phys.Rev.A71.062710}. In comparison, the spin-polarized case $|2\rangle$+$|2\rangle$ shows a reversed ordering and a much larger splitting of about 12 mG~\cite{Chevy2005Phys.Rev.A71.062710, Zhang2022Chin.Phys.B31.063402}.
This predicted 3.6 mG splitting is among one of the smallest known for atomic $p$-wave resonances. Resolving such a feature requires sub-milligauss spectral resolution and magnetic field stability better than $10^{-6}$, presenting a significant experimental challenge.

Here we report a precision measurement of dipolar splitting in a two-component $^6$Li Fermi gas and present two key findings. First, using high-resolution trap-loss spectroscopy combined with active magnetic field stabilization, we resolve a previously unobserved resonance doublet with a splitting of $3.5 \pm 0.1$ mG near 185 G. Our analysis incorporates a two-body loss model that includes all relevant $m_\ell$ components and explicitly accounts for magnetic noise broadening. We find that including magnetic noise is essential: without noise suppression, the doublet structure becomes unresolvable and may be misidentified the sign of the splitting. We also confirm a larger splitting of $11.2 \pm 0.1$ mG in the spin-polarized configuration at 215 G. Both measurements are accurate to within 0.1 mG and show excellent agreement with CC predictions, providing high-precision benchmarks for refining interatomic interaction potentials in $^6$Li\cite{Kempen2004PRA70.050701R}.

Second, we determine the relative positions of the $m_\ell$ components by analyzing the momentum distribution of dissociated molecules formed in different orbital states. The $m_\ell = 0$ resonance exhibits a characteristic double-peak profile along the quantization axis, while the $|m_\ell| = 1$ components display nearly isotropic distribution. These patterns are consistent with the angular nodal structures shown in Fig.~\ref{physics}. Based on these observations, we confirm for the first time that the dipolar splitting observed in the spin-mixture $|1\rangle + |2\rangle$ resonance corresponds to the geometry illustrated in Fig.~\ref{physics}(b), where the $m_\ell = 0$ component lies below the $|m_\ell| = 1$ doublet—opposite to the ordering observed in the spin-polarized $|2\rangle + |2\rangle$ case.

\section{Experiment and Methods}
We begin with a spin-balanced mixture of atoms in the $|1\rangle$ and $|2\rangle$ hyperfine states, confined in a crossed beam optical dipole trap. Following established evaporative cooling protocols~\cite{ Peng2024Commun.Phys.7.101}, we reduce the gas temperature to $T = 0.094\ \mu\text{K}$ at 320 G. Maintaining this ultralow temperature—the lowest achievable in our apparatus—is crucial for minimizing thermal broadening in the atom-loss spectra. The final sample consists of approximately $6 \times 10^4$ atoms per spin component, confined in a harmonic potential with trap frequencies $\omega_x = 2\pi \times 32 \pm 3$ Hz , $\omega_y = 2\pi \times 136 \pm 2$ Hz and $\omega_z = 2\pi \times 138 \pm 2$ Hz, and resulting in $T/T_F \approx 0.42$ ($T_F$ is the Fermi temperature).

To obtain the atom-loss spectrum near the resonances $B_{\text{res}}$, the magnetic field is initially set to $B_i = B_{\text{res}} + 20\ \text{mG}$, where off-resonant interactions are negligible. The field is then ramped to a target value $B_f$ and held for 70 ms to allow resonant collisions. This hold time is optimized to enhance spectroscopic contrast while limiting atom loss to about 50$\%$ of the initial population.
After the interaction, the field is returned to $B_i$, and the remaining atom number is recorded via absorption imaging after 1 ms of ballistic expansion. The gas temperature is extracted simultaneously by fitting the density profile of the expanded cloud.

To meet the sub-milligauss magnetic field resolution and exceptional field stability, we developed a magnetic field control system with long-term stability better than 1 part per million and a field adjustment resolution of 1.3 mG. This performance is achieved through active feedback stabilization and low-noise bipolar current sources, as described in Refs.~\cite{Chen2021Phys.Rev.A103.063311,Liu2023Rev.Sci.Instrum.94.053201, Li2022arXiv:2212.08257}.
In addition to long-term stability, short-term fluctuations arising from ambient 50~Hz line noise posed a major challenge. We identified inductive pickup as the dominant source, introducing root-mean-square (RMS) field noise of 1.3 mG. After implementing magnetic compensation, this noise was suppressed to 0.1 mG, as verified through radio-frequency spectroscopy. Further details on magnetic field calibration, noise measurement and compensation are provided in SM Section II. 

To identify the $m_\ell$ peaks, we measure the dissociate momentum distribution of different $m_\ell$ molecules. We first ramped the magnetic field to the resonance position corresponding to specific $m_\ell$ value, allowing $p$-wave molecules to form with well-defined orbital angular momentum. Then, a 25~$\mu$s resonant light pulse was applied to selectively remove unpaired atoms, while the molecules remained unaffected due to their low Franck–Condon overlap\cite{,Inada2008PRL101.100401}.
Immediately afterward, the magnetic field was rapidly increased above the resonance to dissociate the molecules into free atoms. This process converts the molecular orbital angular momentum into the relative momentum of the atom pairs. To image the resulting distribution, we turn off the optical trap and allow the atoms to ballistically expand for 0.6 ms, then take absorption images in a plane perpendicular to the magnetic field.  Similar dissociation imaging techniques have been demonstrated in $^{40}$K and lattice-confined $^6$Li systems~\cite{Gaebler2007PhysRevLett.98.200403,Waseem2016JPB49.204001}. Experimental details are provided in SM Section~III.

We observe that molecules in the $m_\ell = 0$ state produce a characteristic double-lobed dissociation pattern aligned along the magnetic field axis, reflecting the nodal structure of the $p$-wave orbital. In contrast, the $|m_\ell| = 1$ states yield nearly isotropic, single-peaked momentum distributions. This striking difference provides a clear signature of the orbital symmetry and confirms the identity of the dipolar-split resonance components.

\begin{figure*}[htbp]
	\begin{center}
		\includegraphics[width=0.9\textwidth,angle=0]{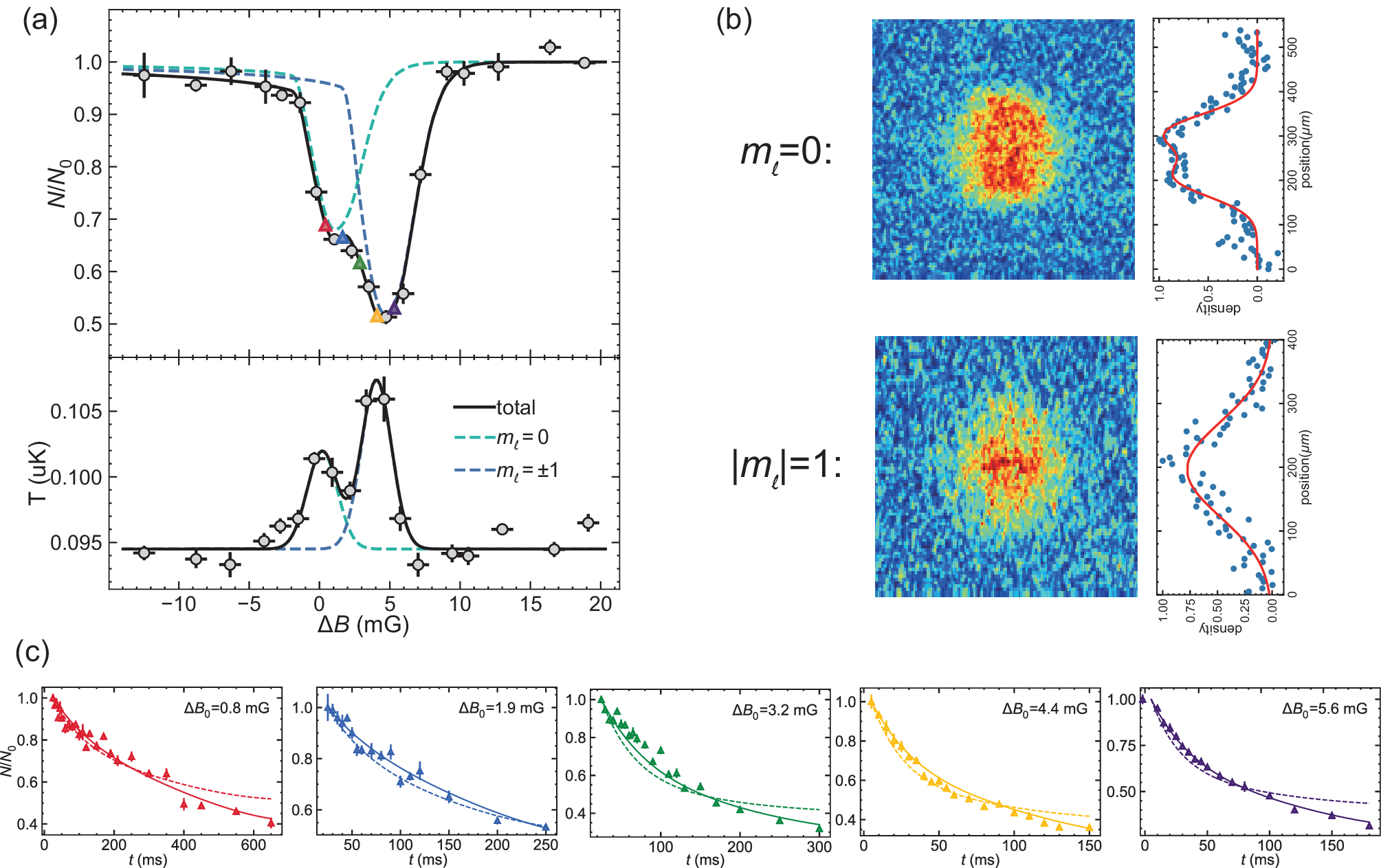}
	\end{center}
\caption{
	(a) Atom-loss and temperature near the $|1\rangle$+$|2\rangle$ \textit{p}-wave FR at 185 G. Black dots are the measured residual atom number (normalized to the initial atom number) with vertical error bars indicating standard deviations from three repetitions and horizontal error bars reflecting the RMS magnetic field noise of 0.7 mG. Solid black curve is a fit of Eq.~\eqref{eq4}, yielding: $\delta_{12}^e = 3.5 \pm 0.1$ mG, and $r_1 = 0.66 \pm 0.01$.  The origin of the horizontal axis corresponds to the fitted $m_\ell = 0$ resonance center at  $B_0 = 185.2175 \pm 0.0002$ G. Blue and cyan dashed curves indicate the $|m_\ell| = 1$ and $m_\ell = 0$ components, respectively. The lower panel shows a double-Gaussian fit to the temperature data, with an initial value $T_0 = 0.094\ \mu$K.	
(b) Column density of atoms dissociated from $m_\ell = 0$ and $|m_\ell| = 1$ molecules, averaged over 8 measurements with approximately $10^4$ molecules. The 1D density along the radial ($R$) direction shows a clear double-peak structure for the $m_\ell = 0$ component.
(c) Time evolution of atom number at selected magnetic detunings, as indicated in (a). Solid and dotted lines represent fits to two-body and three-body decay models, respectively.
}
	\label{loss}
\end{figure*}

\section{Results and Analysis}
We investigate the $|1\rangle$+$|2\rangle$ \textit{p}-wave FR near 185 G using a balanced spin mixture. Figure~\ref{loss}(a) shows the measured magnetic-field-dependent atom-loss spectrum, which exhibits a clear doublet structure.  Figure~\ref{loss}(b) display the dissociation momentum profiles of molecules formed near each peak. The upper panel corresponds to the lower-field peak in Fig.~\ref{loss}(a), and is identified as the $m_\ell = 0$ component based on its characteristic double-lobed structure aligned along the quantization axis—consistent with the nodal pattern of a $p$-wave wavefunction and in agreement with CC predictions\cite{Zhang2022Chin.Phys.B31.063402}. The lower panel, which lacks this double-lobed structure, is assigned to the $|m_\ell| = 1$ component.

A double-peak structure is also observed in the gas temperature (bottom panel of (a)), which increases proportionally with atom loss. For simplicity, we fit the temperature data using a double-Gaussian function, with equal widths constrained across both peaks. The extracted temperature rise associated with the $m_\ell = 0$ component is $0.008 \pm 0.001\ \mu\mathrm{K}$, while that for the $|m_\ell| = 1$ component is $0.013 \pm 0.002\ \mu\mathrm{K}$. The Gaussian width is $1.1 \pm 0.1$ mG, and the peak separation is $3.9 \pm 0.3$ mG\cite{Temperature}. 

To quantify the atom loss, we measure the time evolution of the atom number near the double resonance peaks, as shown in Fig.~\ref{loss}(c). The decay dynamics follow a two-body inelastic collision model:
\begin{equation}
	\frac{dN}{dt} = -\frac{L_2}{V_{\text{eff}}} N^2,
\end{equation}\label{eq:1}
where $V_{\text{eff}} = \sqrt{8} \left( {2\pi k_B T_0}/{m \widetilde{\omega}^2} \right)^{3/2}$ is the effective trap volume with $\widetilde{\omega}=(\omega_x \omega_y \omega_z)^{1/3}$. The fitted $L_2$ at different magnetic detunings $\Delta B_0 = B - B_0$ (where $B_0$ ($B_{\pm 1}$) is the resonance centers for $m_\ell = 0$ ($|m_\ell| = 1$) are $0.50(1)$, $0.93(1)$, $1.61(3)$, $2.53(2)$, and $2.27(3) \times 10^{-16}$ m$^3$/s for detunings of 0.8, 1.9, 3.2, 4.4, and 5.6 mG, respectively.
For comparison, we also fit the data with a three-body loss model (dashed curves in (c)). In all cases, two-body loss clearly dominates. However, near the crossover between the $m_\ell = 0$ and $|m_\ell| = 1$ resonances—such as at $\Delta B_0 = 1.9$ mG (blue)—three-body effects may become non-negligible. 

To accurately model the atom-loss spectrum and extract the peak separation, we begin with the inelastic rate coefficient~\cite{Kurlov2017Phys.Rev.A95.032710,Waseem2017Phys.Rev.A96.062704,Peng2024Phys.Rev.A110.L051301} 
\begin{equation}
	\beta_{\text{in}}(k) = \frac{24\pi \hbar}{m V_1} \frac{k^2}{\left( {1}/{V_p} + k_e k^2 \right)^2 + \left( {1}/{V_1} + k^3 \right)^2}.
\end{equation}
Here, $V_p = V_{\text{bg}} (1 - \Delta / \Delta B)$ is the scattering volume, $V_1$ represents the imaginary part accounting for inelastic loss, and $k_e$ is the effective range. The values of $V_{\text{bg}}$ and $\Delta$ are taken from Ref.~\cite{Fuchs2008Phys.Rev.A77.053616} and are assumed to be the same for all $m_\ell$ resonances: $V_{\text{bg}} = -4.52 \times 10^4\ a_0^3$ and $\Delta = -39.54$ G for the $|1\rangle$+$|2\rangle$ resonance; $V_{\text{bg}} = -4.28 \times 10^4\ a_0^3$ and $\Delta = -25.54$ G for the $|2\rangle$+$|2\rangle$ resonance.
Given the collision energy $E = \hbar^2k^2/m$, the thermally averaged two-body loss rate $L_2$ is calculated by averaging the inelastic rate over a Maxwell-Boltzmann distribution:
\begin{equation}
	L_2 = \frac{2}{\sqrt{\pi}(k_BT)^{3/2}} \int_0^\infty \sqrt{E} e^{-E/k_BT} \beta_{\text{in}}(E) dE.
	\label{TheramlAverage}
\end{equation}

Then, we extend the model to include both $m_\ell = 0$ and $|m_\ell| = 1$ components:
\begin{equation}
	\frac{dN}{dt} = -\kappa \frac{r_1 L_{2}(m_\ell=0) + (1 - r_1)L_{2}(|m_\ell|=1)}{V_{\text{eff}}} N^2,
	\label{eq4}
\end{equation}
where $r_1$ is the fractional contribution from the $|m_\ell| = 1$, and $\kappa$ is a global scaling factor. The dipolar splitting is defined as $\delta_{12}^e = |B_0 - B_{\pm1}|_{12}$.  We incorporate temperature fluctuations and 50 Hz magnetic noise in the fitting model via	$B(t) = B_f + B_{\text{pp}}/2 \sin(100\pi t)$. 

Fitting the data yields a dipolar splitting of $\delta_{12}^e = 3.5 \pm 0.1$ mG. The extracted imaginary scattering volume is $V_1 = (7.6 \pm 5.9) \times 10^{-20}$ m$^3$, and the fitted loss scaling factor is $\kappa = 1.4 \pm 0.2$. The relative strength of the $|m_\ell| = 1$ contribution is $r_1 = 0.66 \pm 0.01$, indicating an approximate 2:1 loss ratio between the $|m_\ell| = 1$ and $m_\ell = 0$ components.

We emphasize that minimizing both gas temperature broadening and magnetic field noise is critical to resolving the small dipolar splitting in the $|1\rangle$+$|2\rangle$ resonance. In SM Section IV, we present data acquired at a higher temperature, where the doublet becomes less distinguishable due to thermal broadening, as described by Eq.~\eqref{TheramlAverage}. Notably, this limitation can be overcome with improved signal-to-noise ratios, and the extracted splitting remains consistent with low-temperature measurements.

In SM Section V, we show measurements taken under elevated 50 Hz magnetic field noise. These results illustrate that uncontrolled magnetic noise significantly distorts the loss spectrum, potentially misplacing the positions of the $m_\ell = 0$ and $|m_\ell| = 1$ peaks and leading to incorrect assignments. This underscores the necessity of active magnetic noise compensation for accurate resonance characterization.

\begin{figure}[htbp]
	\begin{center}
		\includegraphics[width=1\columnwidth, angle=0]{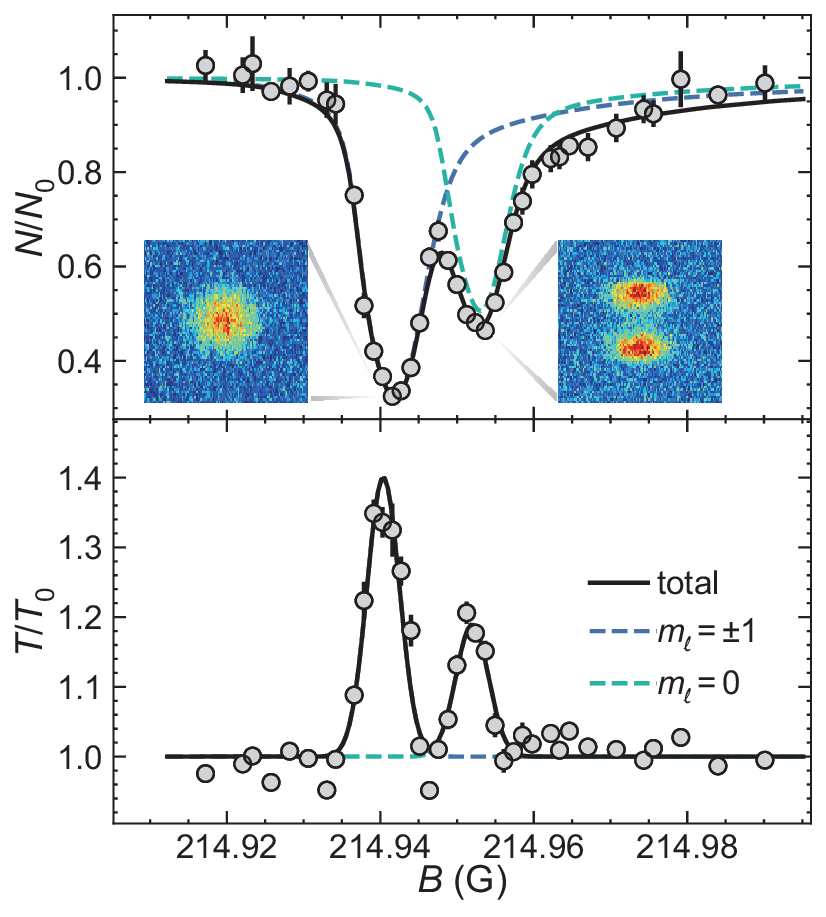}
	\end{center}
\caption{
	Atom-loss spectrum and temperature variation near the $|2\rangle$+$|2\rangle$ \textit{p}-wave FR at 215 G, measured and analyzed using the same protocol as in Fig.~\ref{loss}(a). Horizontal error bars reflect an increased RMS magnetic field noise of 1.6 mG. The inset shows column densities of atoms dissociated from $m_\ell = 0$ and $|m_\ell| = 1$ molecules after a 0.5,ms time of flight, averaged over 8 measurements with approximately $2 \times 10^4$ molecules. The solid black curve is a fit to the two-body loss model, yielding a dipolar splitting of $\delta_{22}^e = 11.2 \pm 0.1$ mG, relative strength $r_1 = 0.68 \pm 0.01$, $B_0 = 214.9273 \pm 0.0001$ G, and imaginary scattering volume $V_1 = (1.7 \pm 0.2)\times 10^{-22}$ m$^3$. The initial temperature was $T_0 = 0.139\ \mu\mathrm{K}$.
}
	\label{22_0.00025U0}
\end{figure}

We perform the same measurement on a spin-polarized $^6$Li gas near the $|2\rangle$+$|2\rangle$ \textit{p}-wave FR at 215 G. The spin-polarized sample is prepared by optically removing $|1\rangle$ atoms using a 100 $\mathrm{\mu}$s resonant laser pulse at an initial field $B_i = 194$ G. The magnetic field is returned to $B_i$ to detect the residual  $|2\rangle$ atoms after 100 ms collision. The resulting atom-loss spectrum and associated temperature variation are shown in Fig.~\ref{22_0.00025U0}. Fitting yields a dipolar splitting of $\delta_{22}^e = 11.2 \pm 0.1$ mG, in good agreement with theoretical prediction. The extracted parameters are $V_1 = (1.7 \pm 0.2) \times 10^{-22}$ m$^3$, $\kappa = 7.8 \pm 0.2$, and $r_1 = 0.68 \pm 0.01$—in excellent equal the obtained value in the spin-mixed case, although both of them deviating from CC predictions that suggest negligible decay in the $m_\ell = -1$ channel~\cite{Chevy2005Phys.Rev.A71.062710,Zhang2022Chin.Phys.B31.063402}. Moreover, unlike the spin-mixed case, the higher-field peak is identified as the $m_\ell = 0$ component, as evidenced by the inset momentum profiles. This assignment is consistent with CC predictions and previous measurements in spin-polarized systems.
The temperature of the gas increases with the magnetic field, following the same trend as the atom loss strength.  

\section{DISCUSSION AND CONCLUSION }
In conclusion, we have successfully observed a small energy splitting of 3.5~mG in $^6$Li \textit{p}-wave FR within the $|1\rangle$+$|2\rangle$ channel by suppressing the influence of 50 Hz magnetic field noise. This splitting value is consistent with calculations from the CC model. We have also developed an inelastic two-body analysis model that accounts for the double structure, which can be widely applied to other systems dominated by two-body inelastic collisions.

Additionally, we observe a spin-dependent dipolar splitting in $^6$Li fermi gas. Unlike in the spin-polarized $|2\rangle$+$|2\rangle$ resonance, the $m_\ell = 0$ resonance peak in the $|1\rangle$+$|2\rangle$ channel appears at a lower magnetic field than the $|m_\ell| = 1$ peak. We further suggest that similar inversion behavior may arise in other nonzero partial-wave scattering, such as \textit{d}-wave resonances \cite{Berninger2013Phys.Rev.A87.032517,Cui2017Phys.Rev.Lett.119.203402,Yao2019Nat.Phys.15.570-576}. 

Studying such ultra small dipolar splitting in $^6$Li Fermi gas is both important and necessary. With its relatively small dipolar splitting, $^6$Li serves as an ideal platform for exploring the interplay between competing \textit{p}-wave pairing channels. This balance is crucial for accessing and controlling quantum phase transitions between different superfluid states, as well as for mapping out the full complexity of the phase diagram\cite{Gurarie2005Phys.Rev.Lett.94.230403,Ho2005PRL94.090402}.

However, an open question remains: the relative strength of the $|m_\ell|=1$ component is measured to be 0.66, approximately twice that of the $m_\ell=0$ component, which deviates from theoretical predictions in which the $m_\ell = -1$ state is non-decaying. This discrepancy may indicate the presence of additional decay mechanisms, such as three-body recombination~\cite{Jona-Lasinio2008Phys.Rev.A77.043611} or many-body correlations and could be a universal feature of all \textit{p}-wave systems, not captured in conventional cc models. Further investigation is needed to distinguish these effects from unresolved $m_\ell$ components or thermal excitations.

\section*{Acknowledgements}
We thanks Bowen Si for his helpful discussion. J.~Li thanks Haoran Zeng, and Linyu Zeng for their preliminary investigation on dipolar splitting.
This work is supported by  NSFC under Grant No.11804406 and No.12174458. J.~Li received supports
from Fundamental Research Funds for Sun Yat-sen University 2023lgbj0 and 24xkjc015. L. Luo received supports from Shenzhen Science and Technology Program JCYJ20220818102003006.

\appendix
\renewcommand{\figurename}{SFig.}
\setcounter{figure}{0}
\section*{Supplemental Material}
\section*{I. Dipole--dipole interaction in $^6$Li $p$-wave Feshbach resonances and Coupled-channel bound-state calculations}
As discussed in the main text, the dipolar splitting observed in $p$-wave Feshbach resonances (FRs) of ultracold $^6$Li gases originates from magnetic dipole--dipole interactions between colliding atoms in a closed-channel molecular state. This interaction is intrinsically anisotropic and depends sensitively on the relative orientation of the atomic magnetic dipole moments. The following are special analysis for $^6$Li $p$-wave FRs.

The magnetic dipole--dipole interaction originates from the coupling between the electron spins of the two atoms and is described by the anisotropic spin--spin Hamiltonian\cite{Ticknor2004Phys.Rev.A69.042712SM}:
\[
H_{\mathrm{ss}} = 2D(r) \left(S_z^2 - \frac{1}{3}\mathbf{S}^2\right),
\]
where \(\mathbf{S} = \mathbf{s}_1 + \mathbf{s}_2\) is the total electronic spin and \(D(r)\) is a radial coupling coefficient. Noted that for the molecule in a singlet states with \(S = 0\), both \(S_z = 0\) and \(\mathbf{S}^2 = 0\), so \(H_{\mathrm{ss}} = 0\). Thus, the dipole--dipole interaction has no effect on \(S=0\) states, and no dipolar splitting occurs between different $m_\ell$ if only the singlet molecular states are coupled in FRs.
Only for molecular states with \(S = 1\), leading to observable energy splittings between orbital projections \(m_\ell = 0, \pm1\) via angular momentum conservation. Moreover, different $m_S=-1,0,1$ will result in different dipolar splitting in parity and strength due to the spatial orientation of the dipoles, which is illustrated in  Fig. 1 of the main text. 

\begin{figure}[htbp]
	\begin{center}
		\includegraphics[width=\columnwidth,angle=0]{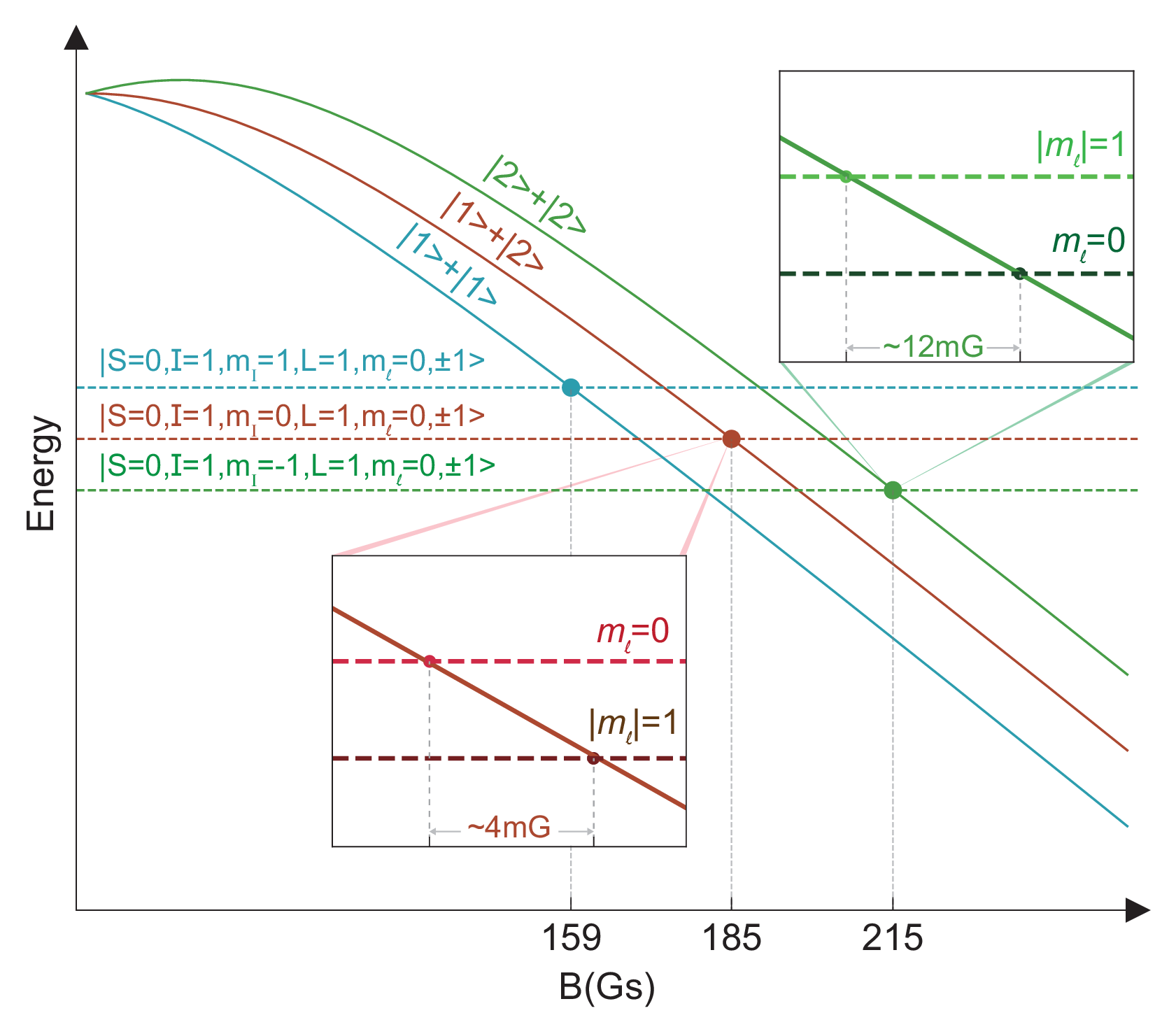}
	\end{center}
	\caption{Energy levels of the atomic and molecular states in $^6$Li $p$-wave Feshbach resonances. Solid lines represent the entrance scattering channels, while dashed lines indicate closed-channel molecular states. The dashed line actually indicates the superposition of several molecular states\cite{Lysebo2009PRA79.062704SM}, which contains a dominated state from singlet channel $S=0$ and several states from triplet $S=1$ channels.  The intersections correspond to the positions of the resonances. Insets show magnified views of the crossings. The order of the crossings for different $m_\ell$ components differs between the \(|1\rangle + |2\rangle\) and \(|2\rangle + |2\rangle\) cases, consistent with the dipolar splitting geometries illustrated in Fig.~1 of the main text.	}
	\label{molecularFig}
\end{figure}

In \textit{p}-wave FRs, entrance-channel are degenerate in $m_\ell$ due to the isotropy of free-space collisions, while closed-channel molecules are bound and experience anisotropic dipole--dipole interactions that split $m_\ell = 0$ and $\pm1$ components, imprinting the structure onto the resonance via Feshbach coupling.
SFigure~\ref{molecularFig} illustrates the schematic energy levels of atomic and molecular states for two $^6$Li atoms in their two lowest hyperfine states near $p$-wave Feshbach resonances. The entrance scattering channels can be expressed in the basis \(|m_{s1}, m_{i1}; m_{s2}, m_{i2}; L\rangle\) as:
\begin{align*}
	|1\rangle + |1\rangle &= \left| -\tfrac{1}{2}, 1;\ -\tfrac{1}{2}, 1;\ L=1 \right\rangle, \\
	|1\rangle + |2\rangle &= \left| -\tfrac{1}{2}, 1;\ -\tfrac{1}{2}, 0;\ L=1 \right\rangle, \\
	|2\rangle + |2\rangle &= \left| -\tfrac{1}{2}, 0;\ -\tfrac{1}{2}, 0;\ L=1 \right\rangle.
\end{align*}

Coupled-channel bound-state calculations using the Fourier Grid Hamiltonian (FGH) method~\cite{Dulieu1995JCP103.60SM} show that each of these entrance channels couples predominantly to a singlet molecular state with total spin \(S=0\), spin projection \(m_S=0\), and nuclear spin \(I=1\), but with different nuclear spin projections: \(m_I = 1, 0, -1\) for the \(|1\rangle+|1\rangle\), \(|1\rangle+|2\rangle\), and \(|2\rangle+|2\rangle\) channels, respectively. These singlet molecular states contribute over 90\% to the total bound-state wavefunction and therefore determine the resonance positions of the respective $p$-wave Feshbach resonances, as shown in SFig.~\ref{molecularFig}.

Because the dipole–dipole interaction Hamiltonian, \(H_{\mathrm{ss}} \), vanishes for \(S = 0\), these dominant singlet states exhibit no dipolar energy shift. The observed dipolar splitting between \(m_\ell = 0\) and \(|m_\ell| = 1\) components instead arises from weak admixture with nearby triplet (\(S=1\)) molecular states. The FGH calculations reveal that the dominant contributing triplet states differ for each resonance: 
\begin{itemize}
	\item \(S=1, m_S=-1, I=2, m_I=2\) for \(|1\rangle+|1\rangle\),
	\item \(S=1, m_S=0, I=0, m_I=0\) for \(|1\rangle+|2\rangle\),
	\item \(S=1, m_S=-1, I=0, m_I=0\) for \(|2\rangle+|2\rangle\).
\end{itemize}
These differences highlight that the effective triplet components responsible for dipolar splitting are distinct between spin-polarized and spin-mixed configurations. 
As shown in the zoomed-in view in SFig.~\ref{molecularFig}, the crossings for different \(m_\ell\) values correspond to the resonant points. In the \( |2\rangle+ |2\rangle \) resonance, the \(m_\ell = \pm1\) resonances appear at lower magnetic fields than the \(m_\ell = 0\) resonance. In contrast, for the \( |1\rangle + |2\rangle \) resonance, the order is reversed. This difference is consistent with the classical picture illustrated in Fig.~1 of the main text.

\section*{II. Magnetic Field Stability and Noise Mitigation}
\subsection*{A. Characterize the magnetic field} 

\begin{figure*}[htbp]
	\begin{center}
		\includegraphics[width=0.8\textwidth]{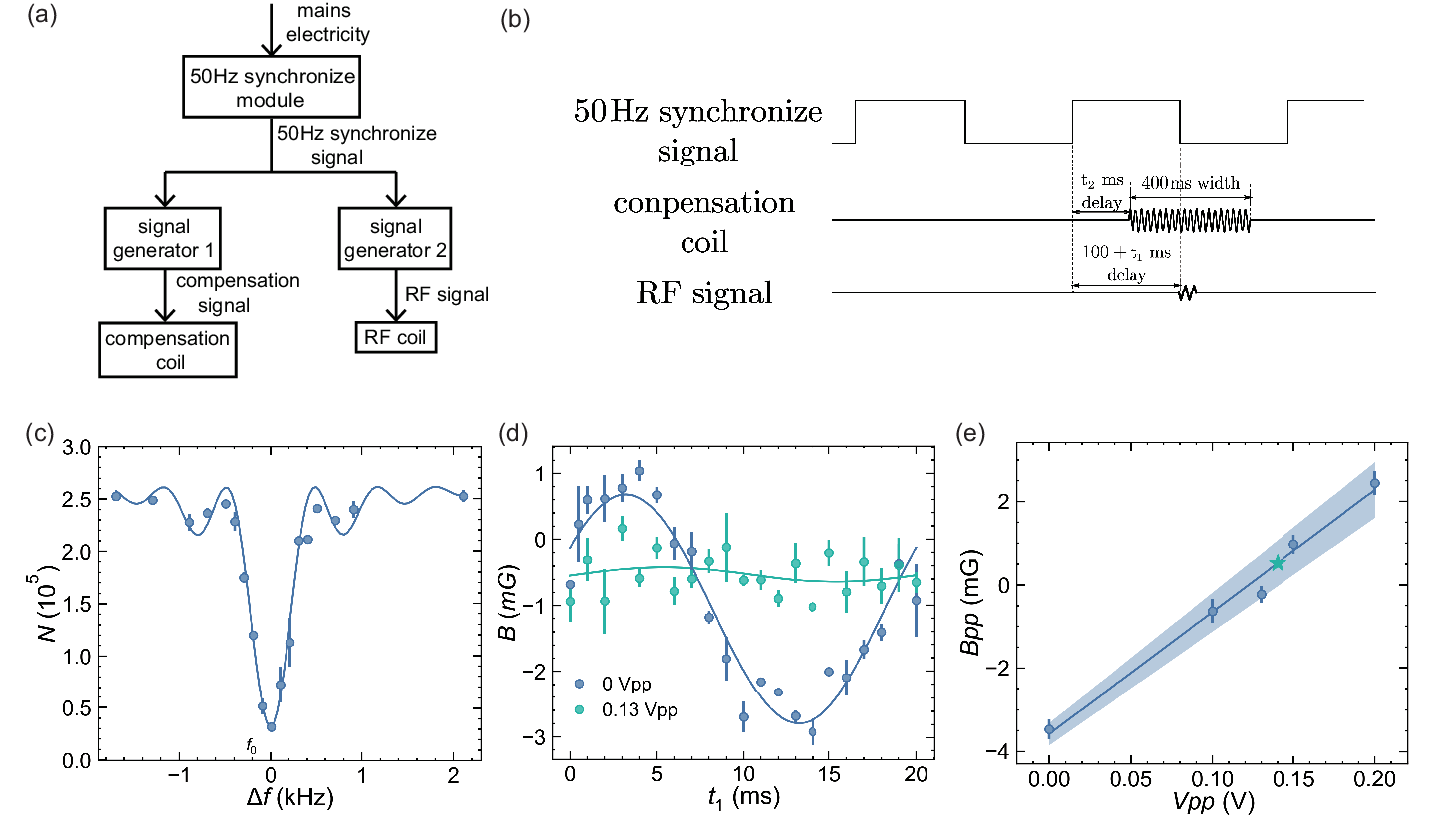}
		\caption{(a) Diagram of 50 Hz magnetic field noise measurement and compensation. The 50 Hz synchronize is generated from the 220 V power supply with a transformer, a 50 Hz band-pass filter and a voltage comparator. (b)Both the compensation coil and RF spectroscopy sequence are triggered by the 50 Hz sync signal, with independently adjustable delays.  The compensation coil introduces a phase-adjustment delay of $t_2$. The RF signal applies a fixed delay of $100\ \mathrm{ms}$ for magnetic field stabilization and a tunable offset $t_1$ vary from 0 to 20 ms. (c) $|1\rangle \rightarrow |2\rangle$ transition at 169 $\mathrm{Gs}$. The transfer rate of atoms from $|1\rangle$ to $|2\rangle$ will fluctuate in response to the magnetic field variations. During the magnetic field noise measurement, we fix the RF frequency at $\Delta f_0 = -100\ \mathrm{Hz}$. We can convert the atom number $N$ to the magnetic field $B$ with the RF spectrum. (d) The solid line in the figure represents the fitted curve for $B = B_p sin(100\pi(t-t_0))+B_0$. The orange line represents the spectrum of a 50 Hz signal with an amplitude of 0.13 $V_{pp}$ from a function generator used for feedback, with a delay $t_2$ = 16 $\mathrm{ms}$. The blue line represents the spectrum without feedback. (e) relationship between the magnitude of a 50 Hz magnetic field signal and the output $V_{pp}$ from the function generator. The positive and negative $B_{pp}$ indicate a phase difference of $180^\circ$.
		}\label{fig:50Hz raw data}
	\end{center}
\end{figure*}

We generate the magnetic field using a pair of current-driven coils, stabilized via a high-precision current transducer and a PID feedback loop that locks the coil-driven current to a reference voltage $U$, as detailed in Ref.\cite{Liu2023Rev.Sci.Instrum.94.053201SM}. The resulting field exhibits a linear dependence on $U$, which we calibrate by measuring the radio-frequency (RF) transition between the $|1\rangle$ and $|2\rangle$ states across a range of fields. A typical result of this RF spectrum is shown in SFig.~\ref{fig:50Hz raw data}(c). Fitting multiple resonance frequencies obtained from the RF spectrum yields the relation $B = (128.8414 \pm 0.0078)\mathrm{G/V} \times U - (0.5897 \pm 0.0178)\mathrm{G}$. The reference voltage is provided by a low-noise source with 10 $\mu$V resolution, corresponding to a magnetic field step size of 1.3 mG, which sets the sampling granularity in Fig.~\ref{loss}.

Although long-term current fluctuations are minimized through active stabilization, residual 50 Hz magnetic field noise originating from the power grid remains the dominant source of short-term field instability. This noise cannot be eliminated by current feedback alone. We characterize it via RF spectroscopy synchronized to the 50 Hz line signal at a background field of 169 G. The resulting sinusoidal field variation, shown in SFig.~\ref{fig:50Hz raw data}(d), has a fitted RMS of $1.3 \pm 0.1$ mG. 

\subsection*{B. Compensate the 50 Hz magnetic field noise}
To counteract 50 Hz magnetic field noise, we employ a compensation coil system in a Helmholtz configuration. The compensation coils are mounted alongside the main coils that generate the primary magnetic field, but with a larger diameter and smaller separation. A function generator is used to drive these compensation coils, producing a synchronized 50 Hz magnetic field for active cancellation.

SFigure~\ref{fig:50Hz raw data}(a) present the diagram of the noise measurement and compensate. Ideally, optimal compensation occurs when the magnetic field generated by the compensation coils is exactly $180^\circ$ out of phase with the ambient 50 Hz noise. However, due to the presence of eddy current effects in the coil structure, we fine-tune the phase by adjusting the turn-on time $t$ of the compensation signal. This allows us to shift the phase of the generated 50 Hz field, as illustrated in SFig.~\ref{fig:50Hz raw data}(b).
The drive current for the compensation coils is small and approximately 2.8 $ \mathrm{mA_{pp}}$, corresponding to an output voltage of 0.14 $\mathrm{V_{pp}}$ from the function generator. 

A typical result of the 50 Hz noise after canceling is presented   in SFig.~\ref{fig:50Hz raw data}(d). The RMS magnetic field fluctuation is reduced to $B_{\mathrm{RMS}} = 0.1 \ \mathrm{mG}$, about 4 times smaller than before. 

\section*{III. Detailed timing sequences of the dissociate momentum distribution measurement of different $m_\ell$ molecules}

\begin{figure}[htbp]
\begin{center}
	\includegraphics[width=0.5\textwidth,angle=0]{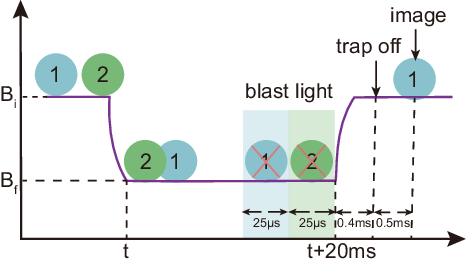}
\end{center}
\caption{Experimental Procedure. Molecules are associated via an exponential magnetic ramp over 20 ms and subsequently dissociated by sweeping back to $B_i$. Resonant light pulses remove unpaired atoms, followed by a ballistic expansion and absorption imaging to resolve the atomic momentum distribution.}
\label{procedure}
\end{figure}

To probe the momentum distribution of $p$-wave Feshbach molecules near the $|1\rangle$+$|2\rangle$ and $|2\rangle$+$|2\rangle$ FR, we prepare ultracold atoms in the \(|1\rangle\) and \(|2\rangle\) hyperfine states (or exclusively \(|2\rangle\) for the 22FR case) at an initial magnetic field \(B_i\) and a temperature of \(T \approx 0.27\,\mu\mathrm{K}\), ensuring that the number of dissociated atoms is sufficient for reliable detection. 

We then ramp the magnetic field from \(B_i\) to the target detuning \(B_f\) over \(20\,\mathrm{ms}\) to associate Feshbach molecules, following an exponential magnetic ramp with a characteristic time constant of $\tau =5 \mathrm{ms}$. After 20 ms of ramping, we apply resonant light pulses—two consecutive 25 $\mu \mathrm{s}$ beams addressing \(|1\rangle\) and \(|2\rangle\) for $|1\rangle$+$|2\rangle$ FR, or a single beam on \(|2\rangle\) for $|2\rangle$+$|2\rangle$ FR—to cleanly remove any unpaired atoms from the trap. 

Next, we rapidly sweep the field back to \(B_i\) to dissociate the molecules into free atom pairs. At \(0.4\,\mathrm{ms}\) after this reversal, when the molecules have fully dissociated, we turn off the optical potential and allow a \(0.5\,\mathrm{ms}\) ballistic expansion. Finally, the atomic momentum distribution is recorded via absorption imaging with a probe beam oriented perpendicular to the magnetic field axis.

\section*{IV. Measurement dipolar splitting with a higher temperature}
		\begin{figure}[htbp]
		\begin{center}
			\includegraphics[width=0.8\columnwidth,angle=0]{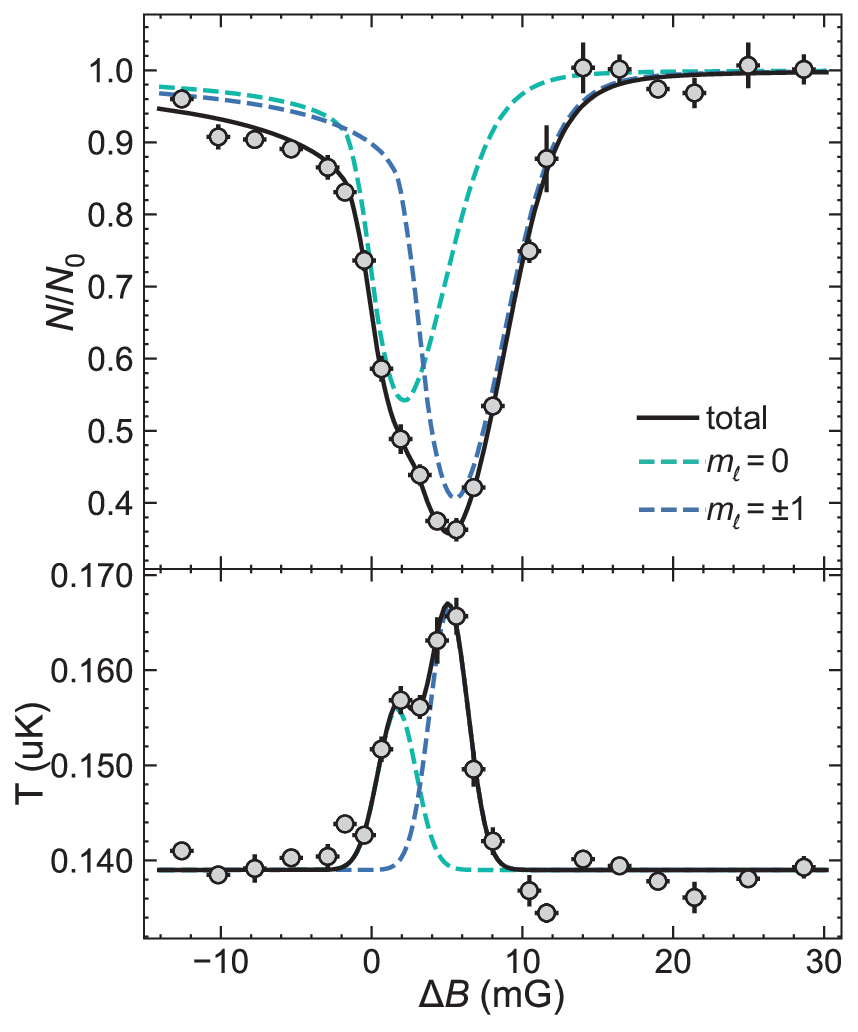}
		\end{center}
		\caption{The same measurement as in Fig. 2 in the main text, but performed at a higher initial temperature of $T_0 = 0.139\ \mu\text{K}$. The fitting yields $\delta_{12}^e = 3.4 \pm 0.2$ mG, $r_1 = 0.63 \pm 0.05$, $\kappa = 0.57 \pm 0.02$, and $V_1 = (0.093 \pm 0.042) \times 10^{-20}\ \text{m}^3$.}	
		\label{12_0.0002U0_Bcom_v2}
	\end{figure}
We further investigate the dipolar splitting of the $|1\rangle$+$|2\rangle$ \textit{p}-wave FR at a higher initial temperature of $T_0$ = 0.139 $\mathrm{\mu}\text{K}$, as shown in SFig.~\ref{12_0.0002U0_Bcom_v2}. Applying the same fitting procedure as in Fig. 2 in the main text, we extract a dipolar splitting of $\delta_{12}^e = 3.4 \pm 0.2$ mG and a relative strength $r_1 = 0.63 \pm 0.05$. These results are consistent with those obtained at lower temperatures and indicate that both the dipolar splitting and the loss ratio between $|m_\ell| = 1$ and $m_\ell = 0$ channels are largely insensitive to the collision energy in this regime.

The fitted loss strength is $\kappa V_1 = (0.57 \pm 0.02)\times (0.09 \pm 0.04) \times 10^{-20} \text{m}^3$, which is approximately one-tenth of the value obtained at lower temperature. This is consistent with the expectation that higher temperatures lead to larger and broader inelastic collision rates near a \textit{p}-wave FR.
Therefore, lowering the temperature is essential for resolving the dipolar splitting more clearly.

	\section*{V. Experiment with larger 50 Hz magnetic field noise}
	
		\begin{figure}[htbp]
		\begin{center}
			\includegraphics[width=0.8\columnwidth,angle=0]{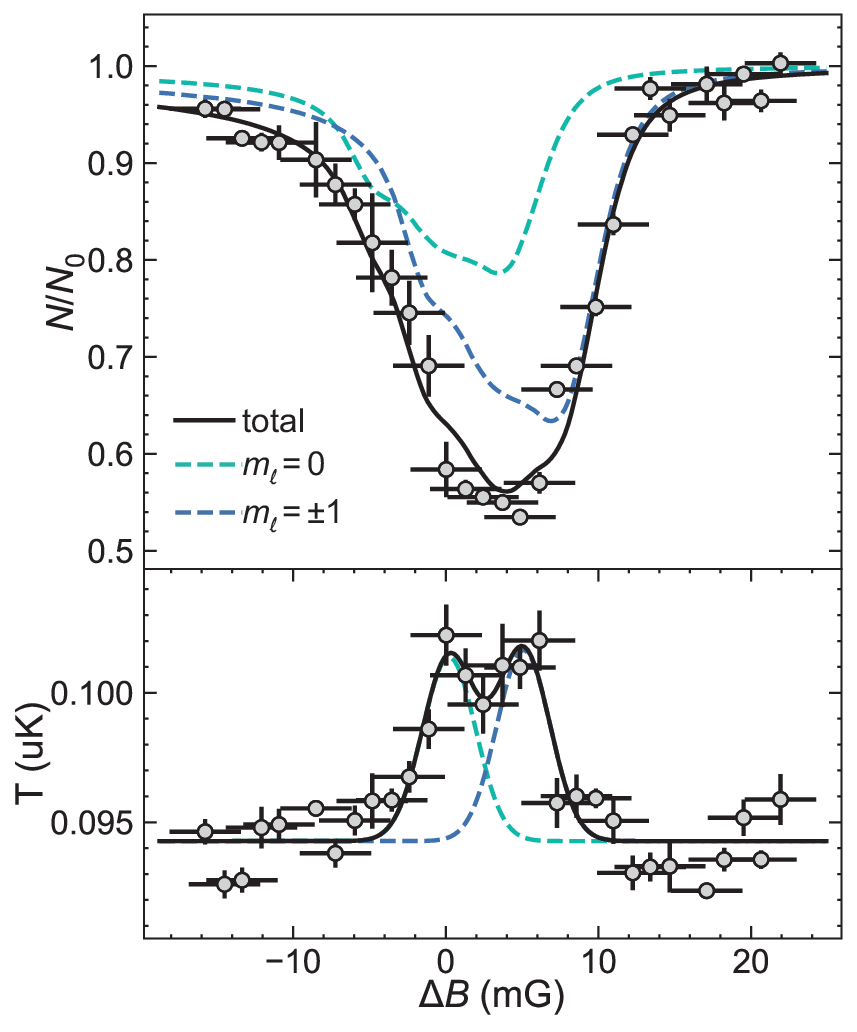}
		\end{center}
		\caption{The same measurement as in Fig. 2 in the main text, but performed with a larger RMS value of 2.4 $\text{mG}$. The dipolar splitting is fixed at $\delta_{12}^e = 3.5\ \text{mG}$, and the relative strength is also set to $r_1 = 0.68$. As shown, the model fit the data very well too, yielding   $\kappa = 1.8 \pm 0.1$, and $V_1 = (0.027 \pm 0.009) \times 10^{-20}\ \text{m}^3$.
		}
		\label{12_0.00015U0}
	\end{figure}
	
	To confirm the suppression of the 50 Hz magnetic field noise is extremely important for detecting the small dipolar splitting.
	As shown in SFig.~\ref{fig:50Hz raw data}(e), by increasing the magnitude or changing the phase of the compensate signal, we can generate a larger 50 Hz magnetic field to simulate the influence. 

	SFig.~\ref{12_0.00015U0} is the result we obtained with a RMS noise of 2.4 $\mathrm{mG}$. It is noted that the atomic loss peaks are significantly broadened by the 50 Hz magnetic field noise. Moreover, the shapes of the peaks are altered and the dipolar splitting structure becomes completely submerged by the noise-induced broadening, making the extraction of the double structure much more difficult. 
	For example, the apparent strength of the double splitting becomes nearly equal by visual inspection, which is inconsistent with the fitting results. This may explain the observation of Ref. ~\cite{Gerken2019Phys.Rev.A100.050701SM}

	\section*{VI. Consideration of the eddy effect in magnetic field}
	Our magnetic field experiences eddy currents due to the water-cooled copper housing of the electromagnetic coils. Using the narrow $s$-wave FR of the $|1\rangle$+$|2\rangle$ channel at 543.4 G, we accurately measured the eddy current time constant in the main magnetic field coils to be $\tau = 5~\mathrm{ms}$~\cite{Chen2021Phys.Rev.A103.063311SM}.
	In this work, we have verified that taking the eddy current effect into account can explain the slight asymmetry observed in the atom-loss spectra shown in SFig.~\ref{12_0.0002U0_Bcom_v2} and ~\ref{12_0.00015U0}, leading to improved fitting results.


\begin{thebibliography}
\bibitem{Temperature}
	\bibinfo{author}{Although the temperature changes across the resonances also show clear visibility and may serve as a viable method to determine the dipolar splitting, we currently lack an accurate model to describe the temperature dynamics ($dT/dt$) due to the complexity of the underlying heating processes. }
\end{thebibliography}
\end{document}